\documentclass[aps,prb,a4paper,10pt,twocolumn,showpacs,floatfix,superscriptaddress,preprintnumbers,longbibliography]{revtex4-2}
\makeatletter

\newcommand{\Rmnum}[1]{\expandafter\@slowromancap\romannumeral #1@}
\makeatother
\usepackage{float}

\usepackage{indentfirst}
\usepackage{dcolumn,graphicx,color,booktabs,microtype,afterpage}
\usepackage{amssymb}
\usepackage{amsmath}
\usepackage[charter,greekuppercase=italicized]{mathdesign}
\usepackage{sidecap}
\usepackage[mathlines]{lineno}
\graphicspath{{./}{figure/}}
\renewcommand{\tablename}{Table}
\makeatletter\renewcommand{\fnum@figure}[1]{\figurename~\thefigure.~}\makeatother
\makeatletter\renewcommand{\fnum@table}[1]{\tablename~\thetable.}\makeatother

\newcount\hh \newcount\mm
\hh=\time \divide\hh by 60
\mm=\hh \multiply\mm by 60 \mm=-\mm
\advance\mm by \time
\def\now{\number\hh:\ifnum\mm<10{}0\fi\number\mm}

\usepackage[colorlinks,plainpages=false,linkcolor=blue,urlcolor=blue,citecolor=blue,pdfpagemode=UseNone,pdfstartview=FitBH]{hyperref}

\newcommand{\tcr}[1]{\textcolor{black}{#1}}

\begin{document}

\makeatletter\renewcommand{\ps@plain}{%
\def\@evenhead{\hfill\itshape\rightmark}%
\def\@oddhead{\itshape\leftmark\hfill}%
\renewcommand{\@evenfoot}{\hfill\small{--~\thepage~--}\hfill}%
\renewcommand{\@oddfoot}{\hfill\small{--~\thepage~--}\hfill}%
}\makeatother\pagestyle{plain}

\preprint{\textit{Preprint: \today, \now}} 
\title{Neutron scattering and muon-spin spectroscopy studies of the magnetic triangular-lattice compounds $A_2$La$_2$NiW$_2$O$_{12}$ ($A$ = Sr, Ba)}
\author{B.\ C.\ Yu}\thanks{These authors contributed equally}
\affiliation{Key Laboratory of Polar Materials and Devices (MOE), School of Physics and Electronic Science, East China Normal University, Shanghai 200241, China}
\author{J.\ Y.\ Yang}\thanks{These authors contributed equally}
\affiliation{Institute of High Energy Physics, Chinese Academy of Sciences (CAS), Beijing 100049, China}
\affiliation{Spallation Neutron Source Science Center (SNSSC), Dongguan 523803, China}
\author{D.\ J.\ Gawryluk}
\affiliation{Laboratory for Multiscale Materials Experiments, Paul Scherrer Institut, CH-5232 Villigen PSI, Switzerland}
\author{Y.\ Xu}
\affiliation{Key Laboratory of Polar Materials and Devices (MOE), School of Physics and Electronic Science, East China Normal University, Shanghai 200241, China}

\author{Q.\ F.\ Zhan}
\affiliation{Key Laboratory of Polar Materials and Devices (MOE), School of Physics and Electronic Science, East China Normal University, Shanghai 200241, China}

\author{T.\ Shiroka}
\affiliation{Laboratory for Muon-Spin Spectroscopy, Paul Scherrer Institut, CH-5232 Villigen PSI, Switzerland}
\affiliation{Laboratorium f\"ur Festk\"orperphysik, ETH Z\"urich, CH-8093 Z\"urich, Switzerland}

\author{T.\ Shang}\email[Corresponding authors:\\]{tshang@phy.ecnu.edu.cn}
\affiliation{Key Laboratory of Polar Materials and Devices (MOE), School of Physics and Electronic Science, East China Normal University, Shanghai 200241, China}
\affiliation{Chongqing Key Laboratory of Precision Optics, Chongqing Institute of East China Normal University, Chongqing 401120, China}

\begin{abstract}
We report on the geometrically frustrated two-dimensional triangular-lattice magnets $A_2$La$_2$NiW$_2$O$_{12}$ ($A$ = Sr, Ba) studied mostly by means of neutron powder diffraction (NPD) and muon-spin rotation and relaxation ({\textmu}SR) techniques. The chemical pressure induced by the Ba-for-Sr substitution suppresses the ferromagnetic (FM) transition from 6.3\,K in the Ba-compound to 4.8\,K in the Sr-compound. 
We find that the $R\bar{3}$ space group reproduces the NPD patterns better than the previously reported $R\bar{3}m$ space group. Both compounds adopt the same magnetic structure with a propagation vector $\boldsymbol{k} = (0, 0, 0)$, in which the Ni$^{2+}$ magnetic moments are aligned ferromagnetically along the $c$-axis. The zero-field {\textmu}SR results reveal two distinct internal
fields (0.31 and 0.10\,T), caused by the long-range ferromagnetic order.
The small transverse muon-spin relaxation rates reflect the homogeneous internal field
distribution in the ordered phase and, thus, further support the simple FM arrangement of the Ni$^{2+}$ moments. The small longitudinal muon-spin relaxation
rates, in both the ferromagnetic- and paramagnetic states of A$_2$La$_2$NiW$_2$O$_{12}$, indicate that spin fluctuations are rather weak.
Our results demonstrate that chemical pressure indeed changes the superexchange interactions in $A_2$La$_2$NiW$_2$O$_{12}$ compounds, with the FM interactions being dominant.  
\end{abstract}


\maketitle\enlargethispage{3pt}

\vspace{-5pt}
\section{\label{sec:Introduction}Introduction}\enlargethispage{8pt}
Geometric frustration occurs when a system of interacting spins is unable
to find its lowest energy state because of how the spins are arranged. 
This property plays an important role at microscopic scales in solids.
In particular, in certain cases, such as in spin glasses, spin ice, and
spin liquids~\cite{binder1986,collins1997,moessner2006,balents2010}, the
localized magnetic moments interact through competing exchange
interactions that cannot be simultaneously satisfied, thus giving rise
to a highly degenerate magnetic ground state. 
For instance, in a spin-liquid system, the constituent spins are
highly correlated, but still strongly fluctuating down to zero
temperature~\cite{binder1986,ma2018,remirez1999,khatua2022,balents2010,shen2018}. 
Such fluctuations lead to remarkable collective phenomena such as emergent gauge fields and fractional excitations~\cite{ramirez1994,greedan2001,balents2010,shen2018}.
Most of the magnetic frustrations have a simple geometric origin~\cite{collins1997,kawamura1998,jaklic2000}, usually occurring in materials
with a 2D triangular- or kagome lattice, or a 3D pyrochlore lattice,
etc., with the nearest-neighbor interactions being antiferromagnetic
(AFM)~\cite{wannier1950,Lv2015}.

\begin{figure}[!htb]
	\centering
	\includegraphics[width=0.49\textwidth,angle=0]{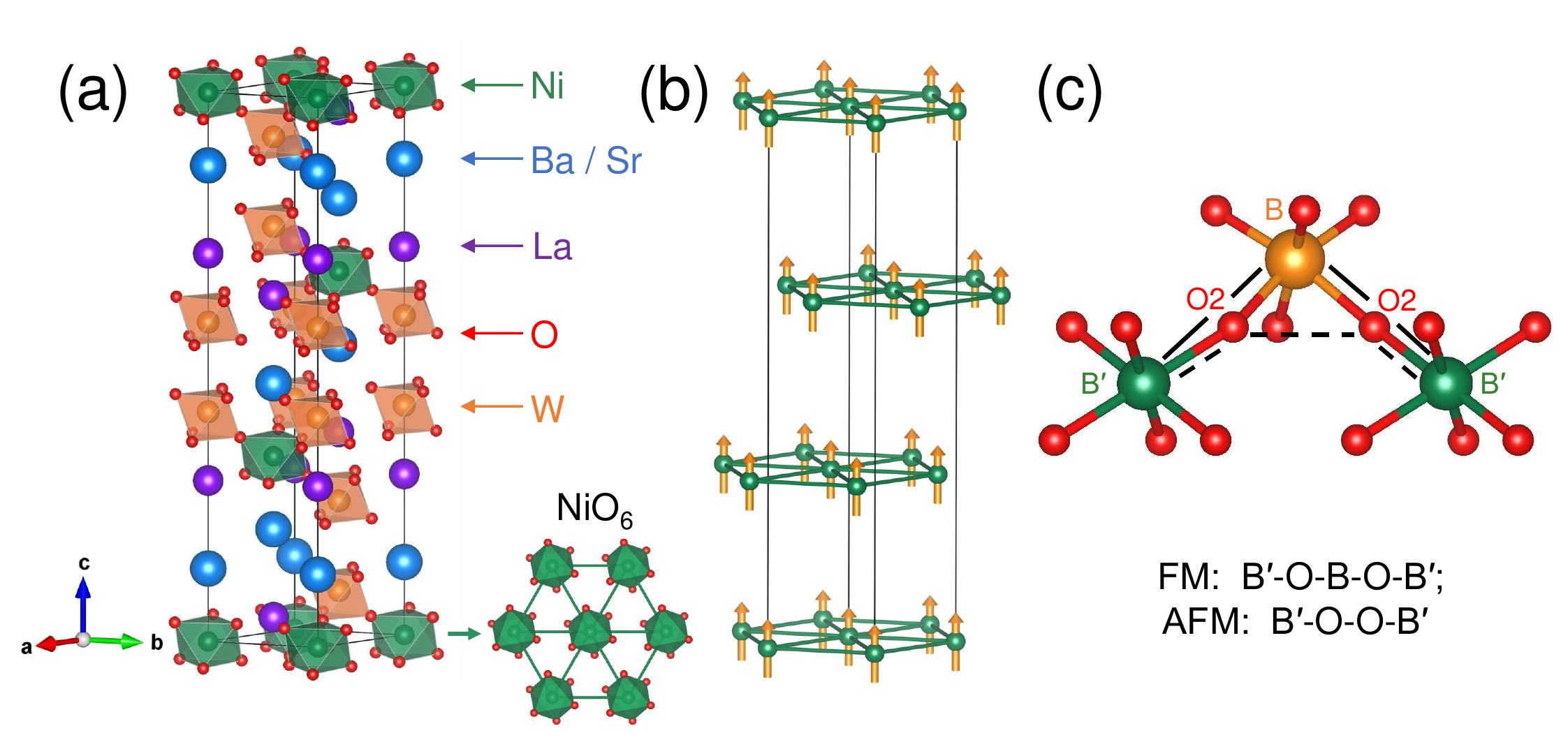}
	\vspace{-2ex}%
	\caption{\label{fig:structure} \tcr{(a) Crystal structure of $A_2$La$_2$NiW$_2$O$_{12}$ ($A$ = Sr, Ba). The Ni layers form
	triangular lattices. 
		(b) Sublattice of Ni$^{2+}$ ions showing that the magnetic
		moments (indicated by arrows) point along the $c$-axis. 
		(c) Pathways of the FM $B$'-O-$B$-O-$B$' (black solid line)
		and AFM $B$'-O-O-$B$' (black dashed line) superexchange interactions.}}
\end{figure}

A two-dimensional triangular lattice with antiferromagnetic interactions
provides one of the prototypes of magnetic frustration~\cite{wannier1950,Lv2015}.
The perovskite-derived compounds $A_4$$B$'$B_2$O$_{12}$ ($A$ = Sr, Ba, La; $B$' = Mn, Co, Ni; $B$ =  Sb, Te, W, Re) represent one such system~\cite{longo1965,rawl2017,saito2019,kojima2018}.
Depending on the valence states of the $B$' and $B$ atoms, the $A$ site can be occupied by either a Sr$^{2+}$ (Ba$^{2+}$) or La$^{3+}$ ion, or by their combinations. 
Here, the magnetic $B$' ions form a layered structure with a 3-fold site symmetry [see Fig.~\ref{fig:structure}(a) for the $B$' = Ni$^{2+}$ case].
Since the magnetic $B$' layers are well separated by the nonmagnetic
$A$- and $B$O$_6$ layers, the former give rise to a magnetic
quasi-2D triangular lattice,
which can potentially host magnetic frustrations. 
 
To date, different magnetic ground states have been found to occur
in the $A_4$$B$'$B_2$O$_{12}$ family~\cite{rawl2017,saito2019,kojima2018},
whose magnetic properties are thought to be determined mostly by the competition between the
ferromagnetic (FM-) $B$'-O-$B$-O-$B$' and antiferromagnetic $B$'-O-O-$B$' superexchange interactions, shown by solid- and dashed lines in Fig.~\ref{fig:structure}(c)~\cite{rawl2017}. The spin state
of the magnetic $B$' ions plays a decisive role in the competition between the
two superexchange interactions. As a consequence, $A_4$Co$B_2$O$_{12}$
(effective spin $S = 1/2$ for Co$^{2+}$) and
Ba$_2$La$_2$NiW$_2$O$_{12}$ ($S$ = 1 for Ni$^{2+}$) are reported to be ferromagnetic, while
Ba$_2$La$_2$MnW$_2$O$_{12}$ ($S$ = 5/2 for Mn$^{2+}$) is reported to be antiferromagnetic~\cite{rawl2017,doi2017}. 
Similar superexchange interactions and their competitions have been observed in other triangular-lattice magnets, e.g., Ba$_3$$B$'Nb$_2$O$_9$~\cite{lee2014,yokota2014,hwang2012,lee2014-2} and $A$Ag$_2$$B$'(VO$_4$)$_2$~\cite{M2012,tsirlin2012}. 
Unsurprisingly, such closely competing interactions can be tuned by either external pressure or by chemical substitution,
each of which able to introduce lattice distortions and to modify
the bond lengths and angles~\cite{M2012,tsirlin2012,sun2019,Sengupta2010,Jiao2020,Gebauer2000}, thus, tuning the magnetic order and frustration.  
For example, in $A_4$Co$B_2$O$_{12}$, the chemical pressure (i.e., the substitution of Ba with Sr and/or La, or W with Re) can tune the FM transition temperature~\cite{rawl2017}.  
However, the effects of chemical pressure on the magnetic properties
of $A_4$Ni$B_2$O$_{12}$ have not been investigated in detail.

To clarify the above issues, in this paper, we synthesized polycrystalline samples of $A_2$La$_2$NiW$_2$O$_{12}$ ($A$ = Sr, Ba)
and studied their magnetic properties by means of mag\-ne\-ti\-za\-tion\mbox{-,} specific heat-, neutron scattering-, and muon-spin rotation and relaxation ({\textmu}SR) measurements. The chemical pressure is introduced by substituting Ba with Sr, which suppresses the 
FM transition temperature from 6.3 down to 4.8\,K, while the magnetic
moments of the Ni$^{2+}$ ions are ferromagnetically aligned along the $c$-axis in both compounds. 
Our results suggest that the chemical pressure indeed changes the superexchange interactions in $A_2$La$_2$NiW$_2$O$_{12}$, with the $B$'-O-$B$-O-$B$' superexchange path dominating
the competition between the FM and AFM interactions. External
pressure on Sr$_2$La$_2$NiW$_2$O$_{12}$ or chemical substitution on the
Ni site may further tune the magnetic interactions and lead
to magnetic frustration.

\section{Experimental details\label{sec:details}}\enlargethispage{8pt} 
The $A_2$La$_2$NiW$_2$O$_{12}$ ($A$ = Sr, Ba) polycrystalline samples were prepared by
the solid-state reaction method. Stoichiometric amounts of La$_2$O$_3$,
BaCO$_3$, SrCO$_3$, NiO, and WO$_3$ powders were used to prepare the
materials. The La$_2$O$_3$ rare-earth oxide was annealed for
15\,hours in atmosphere to remove moisture. The powders were then mixed, ground, and sintered at 1200$^{\circ}$C for 24 hours. After grinding the samples again, the powders were pressed into pellets and sintered at 1200$^{\circ}$C for extra 48 hours. The magnetic-susceptibility and heat-capacity measurements were performed
on a Quantum Design magnetic property measurement system (MPMS) and
physical property measurement system (PPMS), respectively.

Neutron powder diffraction (NPD) measurements were carried out at the Swiss Neutron Source SINQ of the Paul Scherrer Institute in Villigen, Switzerland. The $A_2$La$_2$NiW$_2$O$_{12}$ powder samples were introduced in cylindrical vanadium cans (8\,mm in diameter and 50 mm high) and mounted on a helium cryostat
stick (2--300\,K). High-resolution room-temperature NPD patterns were recorded at the powder diffractometer HRPT [Ge (822), $\lambda = 1.154$~\AA{}].
To discern the magnetic diffraction peaks, high-intensity NPD patterns were collected at 1.7\,K on the DMC diffractometer using a longer wavelength [pyrolitic graphite (002), $\lambda = 2.458$~\AA{}]. 
The collected NPD patterns were analyzed using the Rietveld package of the FullProf suite~\cite{Carvajal1993}.

The bulk {\textmu}SR measurements were carried out at the ge\-ne\-ral\--pur\-pose
surface-muon instrument (GPS) of the Swiss muon source at Paul Scherrer
Institut, Villigen, Switzerland. 
In this study, we performed two types of experiments: zero-field (ZF)-, and longitudinal-field (LF) {\textmu}SR measurements. 
In both cases, we aimed at studying the temperature evolution of the magnetically ordered phase and the spin fluctuations.
The {\textmu}SR spectra were collected upon sample heating and then analyzed by the \texttt{musrfit} software package~\cite{Suter2012}.

\begin{figure*}[htp]
	\centering 
	\includegraphics[width=0.8\linewidth,angle=0]{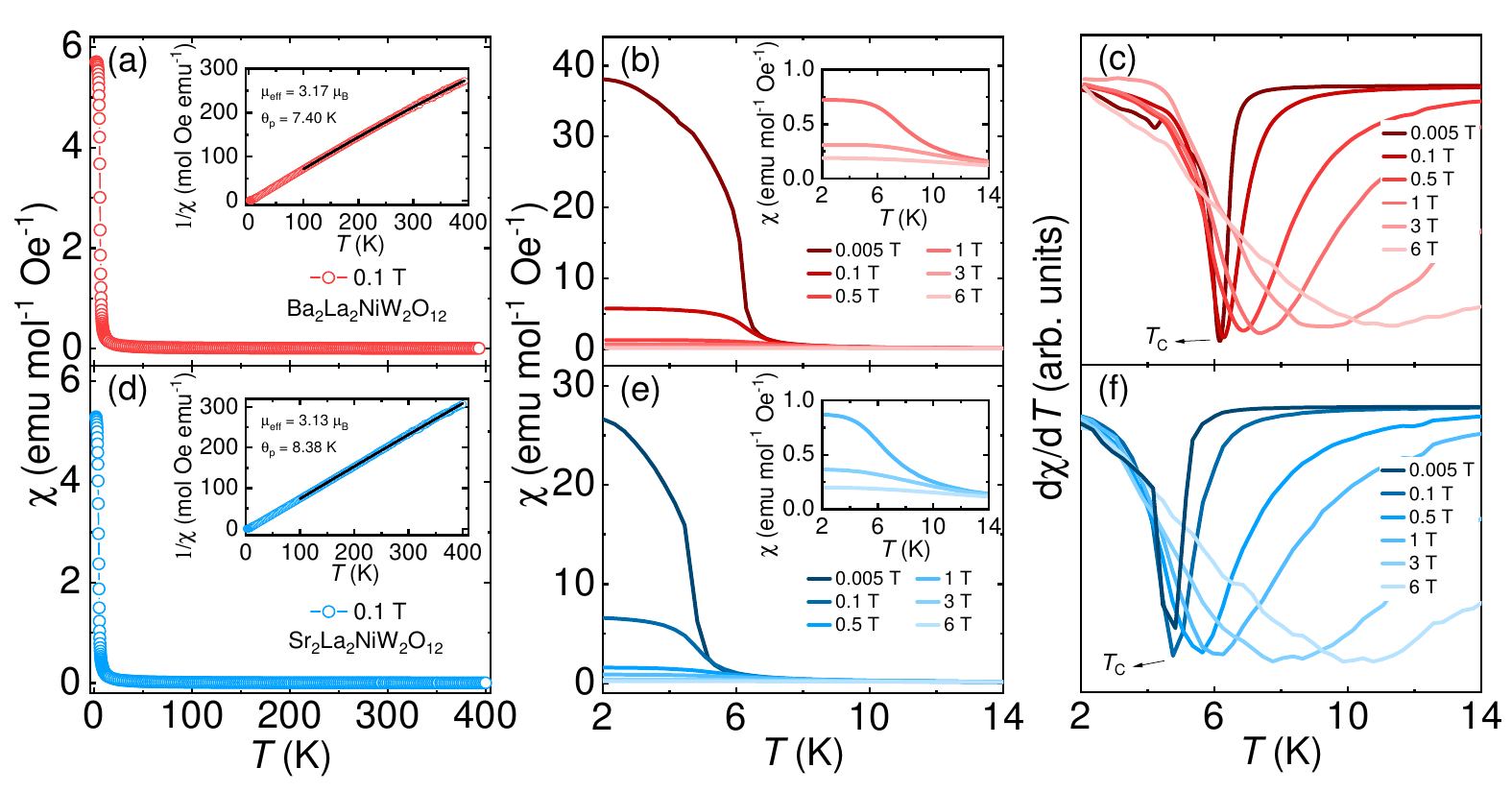} 
	\vspace{-2ex}%
	\caption{\label{fig:chi} (a) Temperature-dependent dc magnetic susceptibility $\chi(T)$ of
		 Ba$_2$La$_2$NiW$_2$O$_{12}$ collected in a field of 0.1\,T. The inset shows the 
		inverse susceptibility  $\chi(T)^{-1}$, with the solid line being a fit to the Curie-Weiss law. (b) Temperature-dependent magnetic susceptibility 
    	of Ba$_2$La$_2$NiW$_2$O$_{12}$ measured in various magnetic fields up to 6\,T. The inset enlarges the $\chi(T)$ curves collected at {\textmu}$_0H = 0.5$, 1, and 6\,T. 
    	Their  derivatives with respect to temperature are shown in panel (c). The temperatures where d$\chi$/d$T$ exhibits a minimum define the Curie temperature $T_c$ and are indicated by an arrow.
    	 The analogous results for Sr$_2$La$_2$NiW$_2$O$_{12}$ are shown in the panels (d)-(f), respectively.}
\end{figure*}

\section{\label{sec:results}Results and discussion}\enlargethispage{8pt} 
\subsection{\label{ssec:chi}Magnetic susceptibility}

\begin{figure}[!htp]
	\centering
	\includegraphics[width=0.45\textwidth,angle=0]{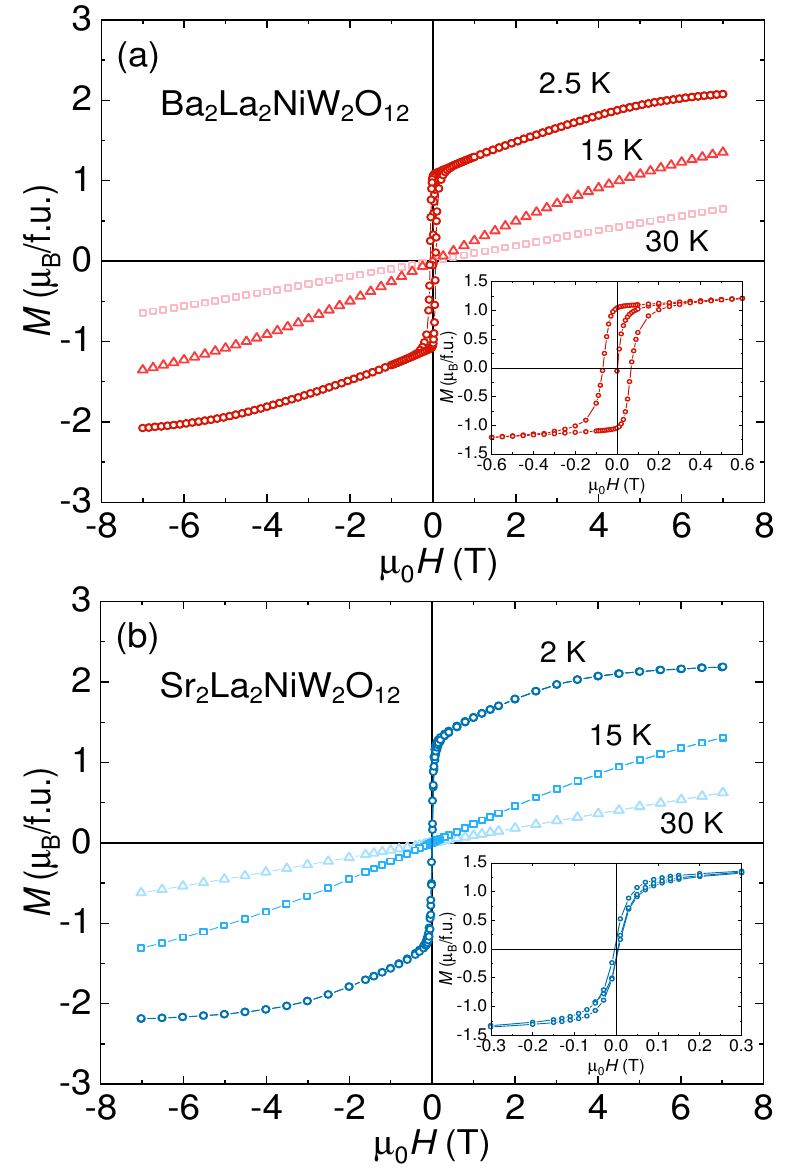}
	\caption{\label{fig:M-H}Field-dependent magnetization $M(H)$ in both the ferromagnetic- and paramagnetic states 
	 of Ba$_2$La$_2$NiW$_2$O$_{12}$ (a) and Sr$_2$La$_2$NiW$_2$O$_{12}$ (b).
    Insets highlight the low-field region of $M(H)$ for Ba$_2$La$_2$NiW$_2$O$_{12}$ (at 2.5\,K) and Sr$_2$La$_2$NiW$_2$O$_{12}$ (at 2\,K),
	clearly showing the hysteresis loops.}
\end{figure} 

\begin{figure*}[!htp]
	\centering
	\includegraphics[width=0.8\textwidth,angle=0]{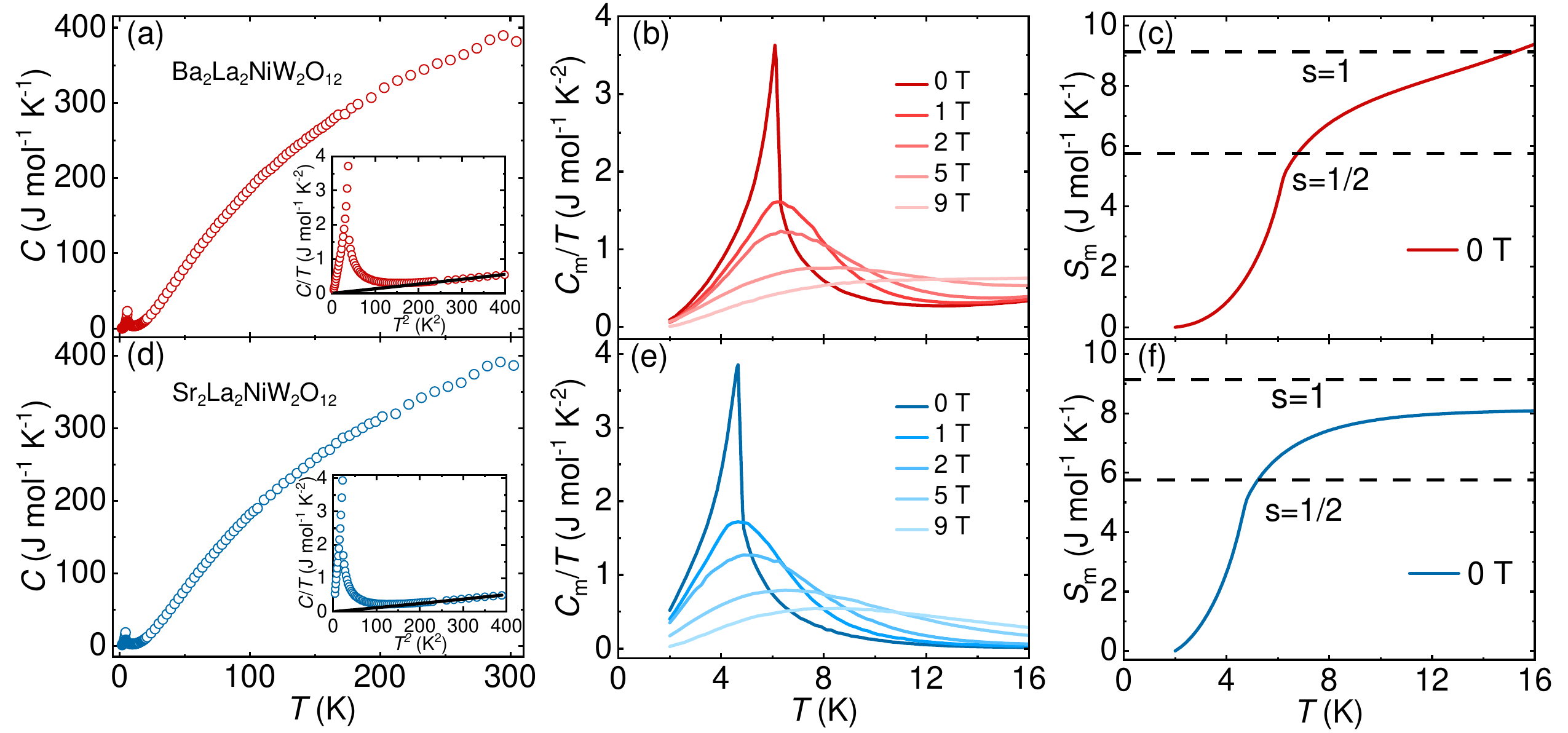}
	\caption{\label{fig:cp}\tcr{(a) Temperature-dependent heat capacity of Ba$_2$La$_2$NiW$_2$O$_{12}$ measured in zero-field condition from 2 to 300\,K. The inset shows the specific heat $C/T$ versus $T^2$ below 20\,K. 
		The solid line is a fit to $C/T$ = $\gamma$ + $\beta$$T^2$ in the paramagnetic state, with $\gamma$ $\equiv$ 0 reflecting the compound's insulating nature. 
     	(b) Temperature dependence of the magnetic contribution to the specific heat $C_\mathrm{m}/T$ for Ba$_2$La$_2$NiW$_2$O$_{12}$ in various magnetic fields up to 9\,T. (c) The zero-field magnetic entropy $S_\mathrm{m}(T)$ obtained from the integration of $C_\mathrm{m}(T)/T$ for Ba$_2$La$_2$NiW$_2$O$_{12}$. The dashed lines mark the entropy values 
        $R\ln(2S + 1)$, with $S = 1/2$ and 1, respectively. The analogous results for Sr$_2$La$_2$NiW$_2$O$_{12}$ are shown in panels (d), (e) and (f), respectively.}}  
\end{figure*}
%

The $A_2$La$_2$NiW$_2$O$_{12}$ samples were first characterized by magnetic-susceptibility measurements. Figures~\ref{fig:chi}(a) and (d) show the temperature-dependent magnetic susceptibility $\chi(T)$ collected in an applied magnetic field of 0.1\,T using a zero-field-cooling (ZFC) protocol. $\chi(T)$ shows a sharp increase close to $T_c$, the
temperature where the Ni$^{2+}$ moments give rise to a FM order. The
Curie temperatures $T_c$ can be determined from the derivative
of susceptibility with respect to temperature d$\chi$/d$T$ [see Fig.~\ref{fig:chi}(c) and (f)] which, in a 0.1-T applied field, provides a $T_c$ of 6.3 and 4.8\,K for
Ba$_2$La$_2$NiW$_2$O$_{12}$ and Sr$_2$La$_2$NiW$_2$O$_{12}$, respectively.
The magnetic susceptibility was also measured under various magnetic fields up to 6\,T. As shown in Fig.~\ref{fig:chi}(b) and (e),
as the magnetic field increases, the transition becomes broader and $T_c$ moves to higher temperatures, both features typical of ferromagnetic materials.   
The insets in Fig.~\ref{fig:chi}(a) and (d) show the Curie-Weiss fits to
the inverse susceptibility (solid lines), which yield a Weiss temperature $\theta_\mathrm{p} = 7.4$\,K
for Ba$_2$La$_2$NiW$_2$O$_{12}$ and $\theta_\mathrm{p} = 8.4$\,K for Sr$_2$La$_2$NiW$_2$O$_{12}$. The positive $\theta_\mathrm{p}$ values indicate that FM interactions are
dominant in both compounds.
The estimated effective moments are $\mu_\mathrm{eff}$ = 3.17\,$\mu_\mathrm{B}$ and 3.13\,$\mu_\mathrm{B}$ for Ba$_2$La$_2$NiW$_2$O$_{12}$ and Sr$_2$La$_2$NiW$_2$O$_{12}$, respectively.
Both are close to the theoretical value of spin-only
Ni$^\mathrm{2+}$ ions (2.83~$\mu_\mathrm{B}$), i.e., assuming a
quenching of the orbital moment, typical of octahedral complexes~\cite{Orchard2003} --- such as the NiO$_{6}$ units in Fig.~\ref{fig:structure}(a).

The FM ground state was further confirmed by field-dependent magnetization
measurements (see Fig.~\ref{fig:M-H}). For $T < T_c$, a small yet clear
magnetic hysteresis loop is observed. For both materials, the
magnetization starts to saturate for $\mu_0H > 5$\,T. After substituting
the Ba with Sr, the magnetism becomes softer. The coercive field
of Ba$_2$La$_2$NiW$_2$O$_{12}$ is about 67\,mT, while, 
in Sr$_2$La$_2$NiW$_2$O$_{12}$, it decreases to 4\,mT.
Thus, in $A_2$La$_2$NiW$_2$O$_{12}$, the chemical pressure suppresses
both the magnetization and the $T_c$, hence suggesting an
enhancement of the magnetic competition. Nevertheless, the FM
interactions remain dominant also in Sr$_2$La$_2$NiW$_2$O$_{12}$. 

\subsection{\label{ssec:cp} Heat capacity}
We measured the zero-field heat-capacity 
of $A_2$La$_2$\-Ni\-W$_2$\-O$_{12}$ from 2 to 300\,K. 
The low-$T$ heat-capacity data were also collected under various 
external fields, up to 9\,T. As shown in Fig.~\ref{fig:cp},
in both compounds, there is a sharp $\lambda$-like transition at
low temperatures, typical of long-range magnetic order.
The $C(T)$ data show a distinct peak at $T_c = 6.1$ and 4.7\,K for Ba$_2$La$_2$NiW$_2$O$_{12}$ and Sr$_2$La$_2$NiW$_2$O$_{12}$, which are consistent with the $T_c$ values determined from magnetization data (see Fig.~\ref{fig:chi}).   
To extract the magnetic contribution, the normal-state (i.e., $T$ $\gg$ $T_c$)
specific-heat data  were fitted to $C/T$ = $\gamma$ + $\beta$$T^2$, where
$\gamma \equiv 0$, due to the insulating nature of both compounds [see solid lines in Fig.~\ref{fig:cp}(a) and (d)]. The derived $\beta$ values are 0.0013  and 0.0012\,J/mol-K$^4$ for Ba$_2$La$_2$NiW$_2$O$_{12}$ and Sr$_2$La$_2$NiW$_2$O$_{12}$, which yield a Debye temperature $\theta_\mathrm{D} = 142$ and 145\,K, respectively. After subtracting the phonon contribution (i.e, the $\beta$$T^2$ term), the magnetic specific heat $C_\mathrm{m}/T$ vs.\ temperature is plotted in Fig.~\ref{fig:cp}(b) and (e) for Ba$_2$La$_2$NiW$_2$O$_{12}$ and Sr$_2$La$_2$NiW$_2$O$_{12}$, respectively.
Upon increasing the magnetic field, the peak at $T_c$ becomes broader
and moves to higher temperatures, once more confirming the FM nature of
the magnetic transition in both materials.
\tcr{The zero-field magnetic entropy $S_\mathrm{m}(T)$ obtained by
integrating $C_\mathrm{m}(T)/T$ is shown in Fig.~\ref{fig:cp}(c)
and (f) for Ba$_2$La$_2$NiW$_2$O$_{12}$ and Sr$_2$La$_2$NiW$_2$O$_{12}$, respectively. 
In both compounds, at temperatures close to $T_c$,
$S_\mathrm{m}$ reaches $R\ln(2)$ (corresponding to $S = 1/2$).
In Ba$_2$La$_2$NiW$_2$O$_{12}$, at temperatures above $T_c$,
$S_\mathrm{m}$ reaches $R\ln(3)$ (corresponding to $S = 1$), while
in Sr$_2$La$_2$NiW$_2$O$_{12}$, $S_\mathrm{m}$ is slightly smaller
than $R\ln(3)$. Such a deviation is most likely 
due to an over-subtraction of the phonon contribution from
the specific-heat data. To properly subtract the phonon contribution
and estimate the magnetic entropy, heat-capacity measurements on
the non-magnetic counterparts, as e.g., $A$$_2$La$_2$ZnW$_2$O$_{12}$, are highly desirable.}


\begin{figure}[htb]
	\centering
	\includegraphics[width=0.49\textwidth,angle=0]{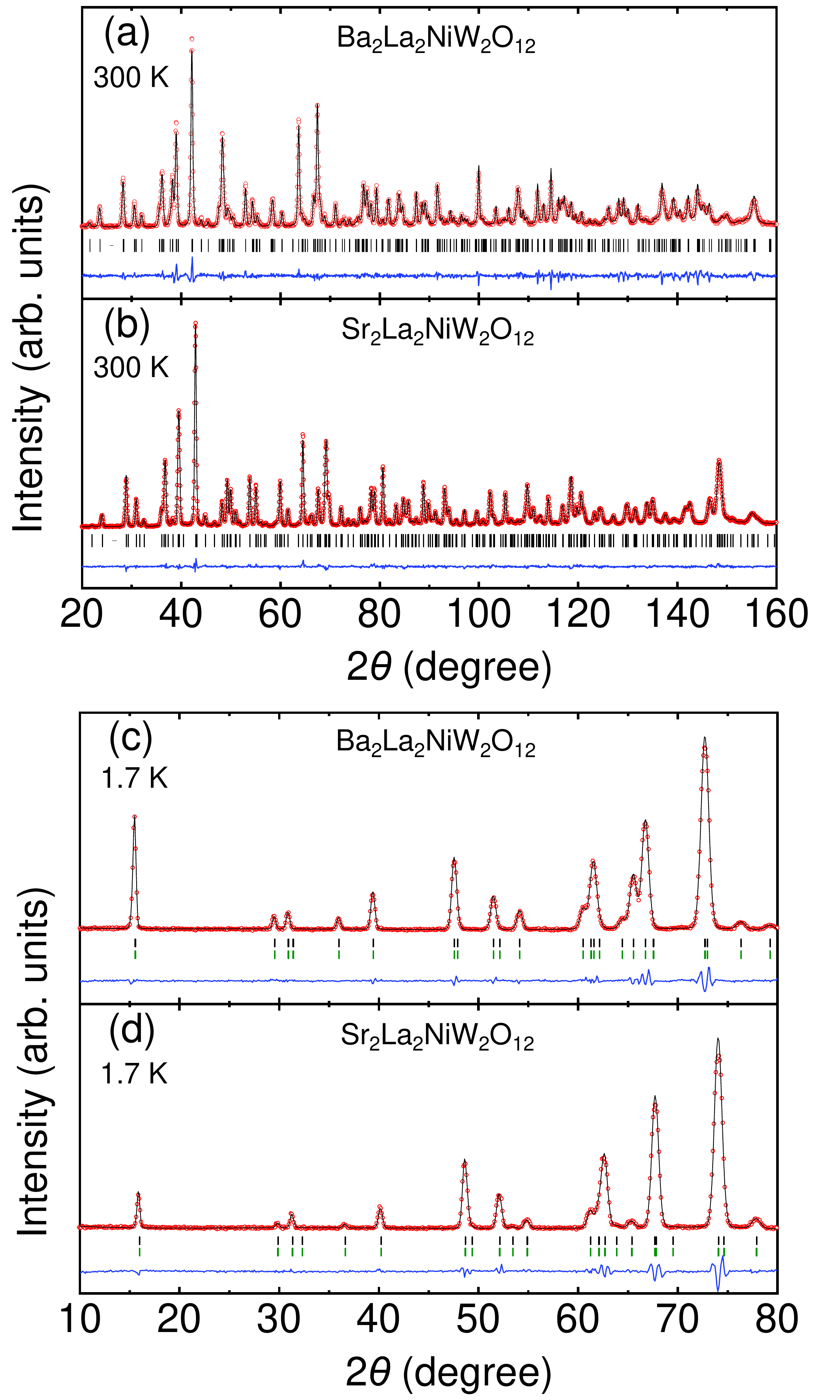}
	\caption{\label{fig:NPD}%
	Rietveld fits of the NPD patterns of Ba$_2$La$_2$NiW$_2$O$_{12}$
	collected in the paramagnetic state (300\,K) (a) and in the
	magnetically-ordered state (1.7\,K) (c). 
	The analogous results for Sr$_2$La$_2$NiW$_2$O$_{12}$ are shown in
	panels (b) and (d), respectively. 
	Red symbols show the experimental data, while black lines are the
	refined profiles. Blue lines at the bottom show the residuals,
	i.e., the difference between the calculated and the experimental data.
	The black and green ticks under the patterns indicate the positions of nuclear and magnetic reflections, respectively.}
\end{figure}
%
%
\subsection{\label{ssec:neutron} Neutron diffraction}

\begin{table*}[!bht]
	\centering
	\caption{\label{tab:atomic} \tcr{Room-temperature
	lattice parameters, atomic positions,
	bond lengths/angles, and goodness of fits for $A_2$La$_2$NiW$_2$O$_{12}$
	($A =$ Ba/Sr)}.
	} 		
	\begin{ruledtabular}
		\begin{tabular}{lccccc}
		\multicolumn{3}{l}{Space group}&\multicolumn{3}{l}{$R\bar{3}$} \\
		\multicolumn{3}{l}{$Z$}&\multicolumn{3}{l}{19} \\
		\multicolumn{3}{l}{$a$(\AA{})}&\multicolumn{3}{l}{5.66126(9)/5.59654(5)} \\
		\multicolumn{3}{l}{$c$(\AA{})}&\multicolumn{3}{l}{27.35363(3)/26.58389(1)} \vspace{3pt}\\ 
		\hline
		$R_\mathrm{p} = 5.53/5.90$\,\%,   \quad	$R_\mathrm{wp}= 7.13/6.42$\,\%,  \quad	$\chi^2_{r} = 2.53/1.97$\, \vspace{3pt} \\ 
		\end{tabular} 
		\\ \vspace{8pt}
		\begin{tabular}{lcccc}
			\textrm{Atom}&
			\textrm{Wyckoff}&
			\textrm{$x$}&
			\textrm{$y$}&
			\textrm{$z$}\\
			\colrule
			Ba/Sr     & $6c$ & 0  & 0  & 0.1329(7)/0.1340(2)  \\
			La        & $6c$ & 0  & 0  & 0.2931(1)/0.2913(2)    \\
			Ni        & $3a$ & 0  & 0  & 0  \\
			W         & $6c$ & 0  & 0  & 0.4182(5) / 0.4215(4)   \\
			O1        & $18f$& 0.4647(5)/0.4445(1)  & 0.4715(8)/0.4472(9)  & 0.1180(3)/0.1216(1)  \\
			O2        & $18f$ & 0.4316(1)/0.4312(6)  & 0.4537(9)/0.4508(6)  & 0.2947(2)/0.2926(2)  \\  \hline 
		     \multicolumn{3}{l}{Bond length: Ni-O2: 2.064(4)~\AA{}/2.051(2)~\AA{}} & 	\multicolumn{2}{l}{Bond length: W-O2: 2.009(6)~\AA{}/2.004(2)~\AA{}} \\
		     \multicolumn{3}{l}{Bond angle: $\angle$Ni-O2-O2: 121.50(5)$^{\circ}$/120.62(4)$^{\circ}$} & \multicolumn{2}{l}{Bond angle: $\angle$O2-W-O2: 84.51(3)$^{\circ}$/84.53(2)$^{\circ}$} \\
		\end{tabular}
	\\ 
\end{ruledtabular}
\end{table*}

To determine the crystal- and magnetic structures of $A_2$La$_2$NiW$_2$O$_{12}$,
neutron powder diffraction patterns were collected at both the
paramagnetic (300\,K)- and ferromagnetic states (1.7\,K).  
The room-temperature patterns were first analyzed by using the space group
$R\bar{3}m$ (No.\ 166), as reported in previous studies~\cite{rawl2017}. 
With this model, the powder x-ray diffraction (XRD) patterns could be fitted reasonably well with a goodness of fit $\chi_{r}^2$ $\sim$ 7.
However, in case of the NPD patterns, although the Bragg peaks were located at the right positions, the $R\bar{3}m$ space group yielded a fairly large $\chi_{r}^2$ $\sim$ 18, 
as evinced also from the clear discrepancy
between the observed- and calculated intensities. This indicates
that the space group {$R\bar{3}m$ does not describe the crystal
structure of $A_2$La$_2$NiW$_2$O$_{12}$ compounds accurately and,
thus, further corrections to the structural model are required. 
Considering that neutron diffraction is more sensitive to the oxygen
atoms than x-ray diffraction~\cite{Furrer2000}, the oxygen positions are 
most likely to require corrections. We found that the space group $R\bar{3}$ (No.~148) reproduces the
NPD patterns quite well. In fact, both $R\bar{3}m$ and
$R\bar{3}$ groups belong to the trigonal system, with the latter
exhibiting slightly different oxygen positions.
Figures~\ref{fig:NPD}(a) and (b) show the Rietveld refinements of NPD
at 300\,K using the $R\bar{3}$ space group for both compounds. 
These refinements yield a significantly reduced $\chi_r^2 \sim 2$,
thus confirming that, in both cases, the $R\bar{3}$ space group
is more appropriate than $R\bar{3}m$. 
With $R\bar{3}$, the NiO$_6$ and WO$_6$ octahedra rotate in
opposite directions around the $c$-axis, which breaks the mirror symmetry.
A similar symmetry breaking has been observed also in the
Ba$_2$La$_2$NiTe$_2$O$_{12}$ compound~\cite{saito2019}. The refined lattice parameters, atomic positions, and bond lengths/angles, 
together with the goodness of fits are summarized in Table~\ref{tab:atomic} for  $A_2$La$_2$NiW$_2$O$_{12}$ compounds.

To clarify the magnetic structure of Ba$_2$La$_2$NiW$_2$O$_{12}$ and Sr$_2$La$_2$NiW$_2$O$_{12}$, the NPD patterns were also collected in the magnetically ordered state
(i.e., 1.7\,K) using long wavelength neutrons ($\lambda = 2.458$~\AA). 
The LeBail fits of the magnetic diffraction patterns
reveal a commensurate magnetic structure with a propagation
vector $\boldsymbol{k} = (0, 0, 0)$ for $A_2$La$_2$NiW$_2$O$_{12}$ compounds. 
For such a magnetic vector, the little group $G_k$ is identical to the
space group $R\bar{3}$ and it includes the symmetry elements
1, 3$^+$, 3$^-$, $\bar{1}$, $\bar{3}^{+}$, and $\bar{3}^{-}$~\cite{Basireps}.  
The magnetic unit cell of $A_2$La$_2$NiW$_2$O$_{12}$ possesses a single orbit with only one site located at the Ni $(0, 0, 0)$ position.
For $\boldsymbol{k} = (0, 0, 0)$, $G_k$ has six different irreducible representations (irreps) $\tau$1, $\tau$2, $\tau$3, $\tau$4, $\tau$5, and $\tau$6, among which only $\tau$1, $\tau$3, and $\tau$5 allow for a long-range magnetic order at the Ni site. Table~\ref{tab:basis} summarizes the basis vectors of $\tau$1, $\tau$3, and $\tau$5 irreps calculated  with BasIreps.
For the $R\bar{3}$ space group, the Ni atoms are located at the 
3$a$ site $(0, 0, 0)$, invariant under all the symmetry operations. 
As a consequence, all the allowed irreps generate a FM coupling with the
spins aligned along the $c$-axis for $\tau$1, or lying within the $ab$-plane for $\tau$3 and $\tau$5 (see details in Table~\ref{tab:basis}).
According to the Rietveld refinements of the 1.7-K NPD pattern [see Fig.~\ref{fig:NPD}(c) and (d)], the best fits were obtained by using the $\tau$1 irrep, yielding the smallest 
$\chi_{r}^2$ = 1.93 and 2.77 for Ba$_2$La$_2$NiW$_2$O$_{12}$ and Sr$_2$La$_2$NiW$_2$O$_{12}$, respectively. The refined magnetic structure is shown in Fig.~\ref{fig:structure}(b). 
The magnetic moments of Ni atoms obtained from the refinements are 1.94(2) and 1.84(3)\,$\mu_\mathrm{B}$ for Ba$_2$La$_2$NiW$_2$O$_{12}$ and Sr$_2$La$_2$NiW$_2$O$_{12}$,
consistent with their saturation magnetization (see Fig.~\ref{fig:M-H}). 


\begin{table}[!bht]
	\centering
	\caption{\label{tab:basis} Basis vectors of irreps $\tau$1, $\tau$3,
	and $\tau$5, as calculated by BasIreps.}
	\vspace{4pt}
	\begin{ruledtabular}
		\begin{tabular}{lccc}
			Site & $\tau$1           & $\tau$3                  & $\tau$5\\ \hline
			Ni   & (0, 0, 1)         & (1, 0, 0)                & (1, 0, 0)\\ 
		         & (0, 0, 0)         & ($-0.58$, $-$1.15, 0)    & (0.58, 1.15, 0)\\
		\end{tabular}
	\end{ruledtabular}
\end{table}

\subsection{\label{ssec:muon} ZF- and LF-{\textmu}SR}  

\begin{figure*}[!hpt]
	\centering
	\includegraphics[width=0.75\textwidth,angle=0]{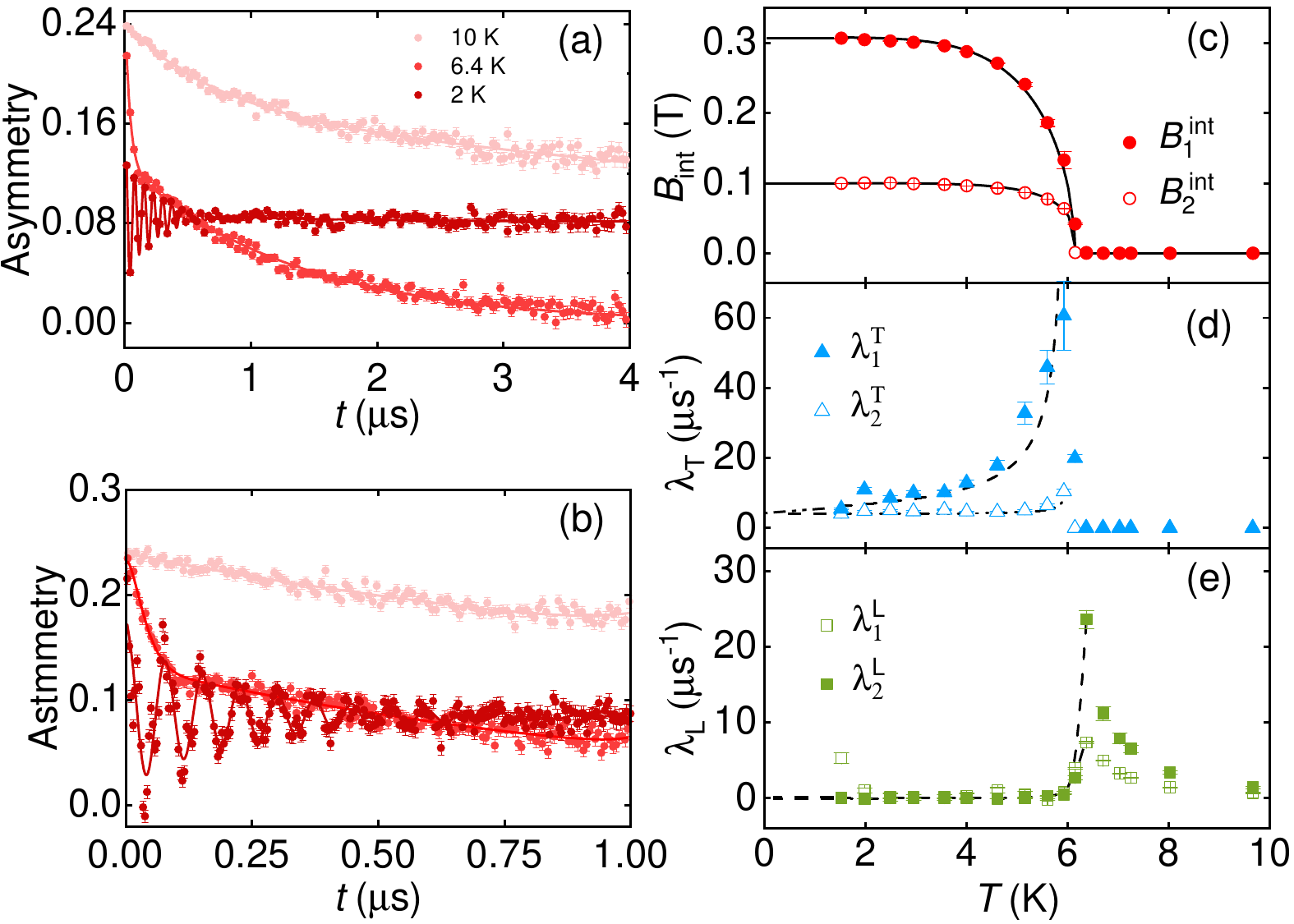}
	\vspace{-2ex}%
	\caption{\label{fig:musr}(a) Representative zero-field {\textmu}SR spectra
	 of Ba$_2$La$_2$NiW$_2$O$_{12}$, collected at various temperatures
	covering both the paramagnetic- and ferromagnetic states. The short-time spectra,
	illustrating the coherent oscillations caused by the long-range FM order,
	are displayed in panel (b). Solid lines through the data are fits to Eq.~\eqref{eq:azc1} and \eqref{eq:azc2} (see text for details).	
	Temperature dependence of the internal field $B_i^\mathrm{int}(T)$ (c), transverse muon-spin relaxation rate (also known as damping rate)
	$\lambda_i^\mathrm{T}$ (d), and longitudinal muon-spin relaxation rate $\lambda_i^\mathrm{L}$ (e) for Ba$_2$La$_2$NiW$_2$O$_{12}$, as derived from the ZF-{\textmu}SR 
	data analysis. Solid lines in (c) are fits to Eq.~\eqref{eq:bint}; dash
	lines in (d) and (e) are guides to the eyes. For clarity reasons, in panel (e), $\lambda_1^\mathrm{L}$ was multiplied by a factor 10.}		
\end{figure*}		

The large gyromagnetic ratio of muons, combined with their
availability as 100$\%$ spin-polarized beams, makes ZF-{\textmu}SR a very sensitive probe for investigating magnetic materials.
Here, to study the magnetic properties of $A _2$La$_2$NiW$_2$O$_{12}$
at a local level, we collected a series of ZF-{\textmu}SR spectra at temperatures covering both the paramagnetic- and ferromagnetic states.
Since neutron diffraction data suggest FM ground states for both Ba$_2$La$_2$NiW$_2$O$_{12}$ and Sr$_2$La$_2$NiW$_2$O$_{12}$ 
(with the Ni$^{2+}$ moments aligned along the $c$-axis),
for our {\textmu}SR measurements we focused on Ba$_2$La$_2$NiW$_2$O$_{12}$
due to its slightly higher $T_c$ value.
In a magnetic material with a long-range order, the time evolution of ZF-{\textmu}SR asymmetry,
$A_\mathrm {ZF}(t)$, encodes both the intrinsic magnetic fields and their distribution at the muon-stopping site~\cite{2011MuonSR}. 
The ZF-{\textmu}SR spectra of Ba$_2$La$_2$NiW$_2$O$_{12}$ collected at different temperatures are shown in Fig.~\ref{fig:musr}(a).
In the paramagnetic state ($T > T_c$), the ZF-{\textmu}SR spectra exhibit a relatively
slow muon-spin depolarization ($\sim$0.5--1\,{\textmu}s$^{-1}$ at 10\,K), indicating rather weak spin fluctuations. 
Considering the two muon-stopping sites in Ba$_2$La$_2$NiW$_2$O$_{12}$, attributed to two distinct oxygen sites (see Table~\ref{tab:atomic}),
the ZF-{\textmu}SR spectra in the paramagnetic state were analyzed using the following model: 
\begin{equation}
\label{eq:azc1}
A_\mathrm{ZF}(t)= \sum\limits_{i=1}^2 A_i e^{-\lambda^\mathrm{L}_it}.
\end{equation}
Here, $\lambda^\mathrm{L}_i$ represent the longitudinal muon-spin relaxation rates,
while $A_i$ are the asymmetries of the two nonequivalent muon-stopping sites.

In the FM state ($T < T_c$), the ZF-{\textmu}SR spectra are characterized by
highly-damped oscillations, typical of long-range magnetic order. 
These are clearly visible in Fig.~\ref{fig:musr}(b), where short-time
oscillations are superimposed on a long-time slow relaxation.
The ZF-{\textmu}SR spectra in the FM state were, hence, analyzed using
the following model:
\begin{equation}
\label{eq:azc2}
A_\mathrm{ZF}(t)= \sum\limits_{i=1}^2A_i[\alpha\cos(\omega_it+\phi)e^{-\lambda^\mathrm{T}_it} + (1-\alpha)e^{-\lambda^\mathrm{L}_it}].
%
%
\end{equation}
Here, $\alpha$ and 1--$\alpha$ are the oscillating (i.e., transverse) and nonoscillating (i.e., longitudinal) fractions of the {\textmu}SR signal, respectively, 
whose initial total asymmetry is equal to $A_1$ and $A_2$. 
In polycrystalline materials with a long-range magnetic order, one expects $\alpha = 2/3$, since statistically one third of the muon spins are aligned parallel to the local field direction 
(i.e., $S_\mu \parallel B_\mathrm{int}$)  and, hence, do not precess; 
$\omega_i$ ($=\gamma_\mu B_i^\mathrm{int}$) represents the muon-spin precession frequency, 
with $\gamma_\mu= 2\pi\times135.5$\,MHz/T the muon gyromagnetic ratio and $B_i^\mathrm{int}$ the local field sensed by muons; 
$\lambda^\mathrm{T}_i$ are the transverse muon-spin relaxation rates, reflecting the internal field distributions; $\phi$ is a shared initial phase.

The derived fitting parameters are summarized in Fig.~\ref{fig:musr}(c)-(e). 
The $B_i^\mathrm{int}$, $\lambda^\mathrm{T}_i$, and $\lambda^\mathrm{L}_i$
all show a distinct anomaly at $T_c$.
The $T_c$ determined from ZF-{\textmu}SR is consistent with the value determined from magnetic susceptibility and heat capacity (see Figs.~\ref{fig:chi} and \ref{fig:cp}). 
As shown in Fig.~\ref{fig:musr}(c), below $T_c$, there are two distinct internal fields, here reflecting the two different muon-stopping sites. 
In the FM state, the temperature evolution of $B^\mathrm{int}_i(T)$
resembles the typical mean-field curve. To estimate the zero-temperature internal field, 
$B^\mathrm{int}_i(T)$ was analyzed by means of
a phenomenological model:
\begin{equation}
	\label{eq:bint}
	B^\mathrm{int}_i(T) = B^\mathrm{int}_i(0) \left[1-\left(\frac{T}{T_c}\right)^{\gamma}\right]^{\delta},
\end{equation}
where $B^\mathrm{int}_i(0)$ is the zero-temperature internal field,
while $\gamma$ and $\delta$ represent two empirical parameters.
As shown by solid lines in Fig.~\ref{fig:musr}(c), the above model describes the data reasonably well, yielding $B^\mathrm{int}_1(0)$ = 0.30\,T
and $B^\mathrm{int}_2(0)$ = 0.10\,T
for Ba$_2$La$_2$NiW$_2$O$_{12}$. \tcr{The resulting
power exponents are $\gamma$ = 5.5(2) and $\delta$ = 0.54(2) for $B_1^\mathrm{int}(T)$, and $\gamma$ = 4.6(2) and $\delta$ = 0.26(1) for $B_2^\mathrm{int}(T)$, respectively.}
The lack of any anomalies in
$B^\mathrm{int}_i(T)$ below $T_c$ is consistent with the simple FM structure of Ba$_2$La$_2$NiW$_2$O$_{12}$ (see Fig.~\ref{fig:structure}). 
In fact, in some complex magnetic materials with multiple transitions,
one observes a more complex $B^\mathrm{int}(T)$,} since changes in
magnetic structure are reflected in the local-field distribution~\cite{Zhu2022}.  
 
\begin{figure}[!htb]
	\centering
	\includegraphics[width=0.48\textwidth,angle=0]{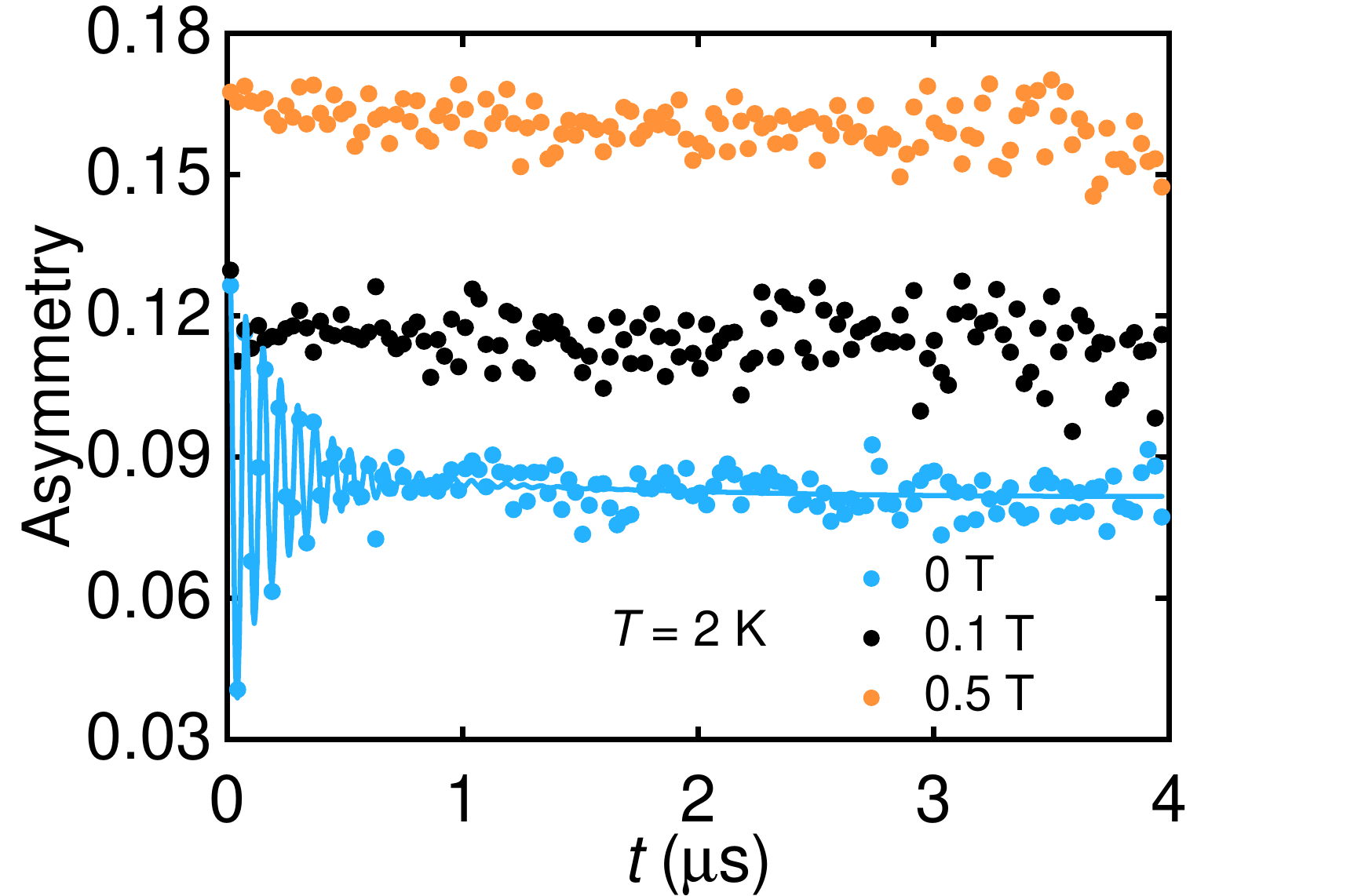}
	\vspace{-4ex}%
	\caption{\label{fig:musrlf}LF-{\textmu}SR spectra of Ba$_2$La$_2$NiW$_2$O$_{12}$
	collected at 2\,K in a magnetic field of 0, 0.1, and 0.5\,T.
    Here, we use a longitudinal muon-spin configuration, i.e.,
	$p_\mu \parallel S_\mu$, with the applied field being parallel
	to the muon-spin direction.	Muon spins are fully decoupled once the external field overcomes the internal field.}
\end{figure}

The transverse muon-spin relaxation rate $\lambda^\mathrm{T}$ reflects the static magnetic field distribution at the muon-stopping site and is also affected by dynamical effects such as spin fluctuations, 
while its longitudinal counterpart $\lambda^\mathrm{T}$ is solely determined by spin fluctuations.
The $\lambda_i^\mathrm{T}(T)$ of Ba$_2$La$_2$NiW$_2$O$_{12}$ exhibits the typical behavior of
magnetic materials with a long-range order~\cite{tran2018,Zhu2022}, i.e., diverging at $T_c$ and continuously decreasing well
inside the magnetic state [see Fig.~\ref{fig:musr}(d)].
In the paramagnetic state, $\lambda_i^\mathrm{T}$ is zero, due to the lack of a magnetic moment in the absence of an external field. 
The $\lambda_i^\mathrm{L}(T)$ in Fig.~\ref{fig:musr}(e) shows a similar
behavior to the $\lambda_i^\mathrm{T}(T)$, i.e., $\lambda_i^\mathrm{L}(T)$
diverges near $T_c$, followed by a significant drop at $T < T_c$,
indicating that spin fluctuations are the strongest close to the onset
of the FM order. Note that, the absolute values of longitudinal relaxation are much smaller than the transverse ones.
Thus, at 1.5\,K, $\lambda^\mathrm{L}$/$\lambda^\mathrm{T} \sim 0.097$
and 0.002 for the two different muon-stopping sites. 
In the paramagnetic state (i.e., $T > 8$\,K), $\lambda_i^\mathrm{L}$ is
also very small, suggesting weak spin fluctuations in both the ferromagnetic and
paramagnetic states of Ba$_2$La$_2$NiW$_2$O$_{12}$. 
Such weak spin fluctuations are further supported by LF-{\textmu}SR measurements.
Figure~\ref{fig:musrlf} shows the 2-K LF-{\textmu}SR spectra
collected in a longitudinal field of 0.1 and 0.5\,T. Once the external
field exceeds the internal field (here, $\sim 0.3$\,T), the {\textmu}SR spectra become
almost flat. This suggests that, in Ba$_2$La$_2$NiW$_2$O$_{12}$,
muon spins are fully decoupled from the electronic magnetic moments
in a field of 0.5\,T. 

\section{\label{ssec:diss} Discussion}   
Although our comprehensive set of measurements suggest that
both Ba$_2$La$_2$NiW$_2$O$_{12}$ and Sr$_2$La$_2$NiW$_2$O$_{12}$ 
have FM ground states, the magnetic susceptibility and neutron diffraction
results indicate that \tcr{the competition between FM- and AFM couplings is indeed tuned}
by the chemical
pressure induced by the substitution of Ba- with the smaller Sr ions.
To understand this, we examine the crystal-structure parameters
of $A_2$La$_2$NiW$_2$O$_{12}$ (see details in Table~\ref{tab:atomic}),
including the bond lengths and angles. The latter are directly
related to the magnetic superexchange interactions and, thus, control the
magnetic properties. In $A_4$$B$'$B_2$O$_{12}$, the $B$'O$_{6}$ octahedra
share their corners with the $B$O$_{6}$ octahedra via oxygen atoms, thus leading to two superexchange interaction paths, i.e.,
$B$'-O-$B$-O-$B$' and $B$'-O-O-$B$' [see details in Fig.~\ref{fig:structure}(c)].
\tcr{According to the Goodenough-Kanamori rule,
which provides the signs of the competitive interactions that are
responsible for non-collinear spin ordering~\cite{Goodenough1955,kanamori1959,Coey2010},
the $B$'-O-$B$-O-$B$' superexchange interaction (with $\angle$O-B-O $\sim$ 90$^\circ$) favors a FM coupling, while the $B$'-O-O-$B$' path (with $\angle$$B$'-O-O $\sim$ 120-180$^\circ$) allows for an AFM coupling. Although the $R\bar{3}$ space group implies
reduced O-B-O and $B$'-O-O bond angles with respect to the previously
reported $R\bar{3}m$ space group~\cite{rawl2017}, the change is such
that the FM or AFM character of the superexchange interactions is maintained.
For instance, in Ba$_2$La$_2$NiW$_2$O$_{12}$, $R\bar{3}m$ gives
$\angle$Ni-O2-O2 = 137.2$^\circ$ and $\angle$O2-W-O2 = 86.7$^\circ$; while
in $R\bar{3}$, these bond angles become 121.5$^\circ$ and 84.5$^\circ$.
Consequently, the $B$'-O-$B$-O-$B$' and $B$'-O-O-$B$' superexchange
interaction paths remain 
valid also in the $R\bar{3}$ space group.}

The competition between these FM and AFM interactions eventually determines
the magnetic ground state of $A_4$$B$'$B_2$O$_{12}$. 
Since Sr has a smaller atomic radius than Ba, by
replacing Ba with Sr, the lattice constants along both the
$a$- and $c$-axis are reduced by a factor of 1.14 and 2.81\%, 
the Ni-O bond length decreases from 2.064~\AA{} to 2.051~\AA{}, while
the Ni-O2-O2 bond angle increases from 121.50$^{\circ}$ to 120.62$^{\circ}$.
By contrast, the W-O bond length and the O2-W-O2 bond angle are
less affected, most likely because the W-O2 layer is further
away from the Ba- or Sr-layers [see Fig.~\ref{fig:structure}(a)].
The O2-W-O2 bond angle increases slightly from 84.51$^{\circ}$ to
84.53$^{\circ}$. \tcr{The changes of Ni-O2-O2 and O2-W-O2 bond angles induced by chemical pressure (i.e., the substitution of Ba by Sr)
tune the competition between FM- and AFM superexchange interactions in $A_2$La$_2$NiW$_2$O$_{12}$.}
The physical pressure might further
tune the competition between the FM- and AFM interactions, and yield
magnetic frustration. 
Previous studies reveal that the magnetic ground states of
$A_4$$B$'$B_2$O$_{12}$ can also be tuned 
by chemical substitution on the $B$ sites~\cite{rawl2017}. 
The substitution on the $B$'-site of Ni may enhance the $B$'-O-O-$B$'
AFM interactions and stabilize the AFM ground state. For instance,
Ba$_2$La$_2$MnW$_2$O$_{12}$ shows an AFM order below 1.7~K~\cite{rawl2017}. 
The Ni$^{2+}$ ions can also be substituted by Cu$^{2+}$ ions, but the
latter case is not yet studied, although it may represent
another interesting compound to exhibit magnetic frustration. Finally, the introduction of magnetic ions on the $A$ site
(e.g., the substitution of Ba$^{2+}$ or Sr$^{2+}$ with Eu$^{2+}$),
whose magnetic interactions can compete with the above superexchange
interactions, may lead to exotic magnetic properties. 

\section{\label{ssec:Sum}Conclusion}
To summarize, we studied the effects of chemical pressure on the
magnetic triangular-lattice compounds $A_2$La$_2$NiW$_2$O$_{12}$ ($A$ = Sr, Ba). 
Their magnetic properties (due to the Ni$^{2+}$ ions) were investigated by means of magnetic susceptibility, specific heat, neutron diffraction, 
and {\textmu}SR spectroscopy. When replacing Ba with Sr, chemical pressure is introduced which can tune the competition between the FM- and AFM superexchange interactions. While the Curie temperature $T_c$ is suppressed
from 6.3\,K to 4.8\,K, the FM interactions still persist
in Sr$_2$La$_2$NiW$_2$O$_{12}$. According to the refinements of neutron
diffraction patterns, in both compounds, the magnetic moments of
Ni atoms are aligned along the $c$-axis, with a propagation vector
$\boldsymbol{k} = (0, 0, 0)$.
By using ZF-{\textmu}SR measurements, we could follow the temperature
evolution of the spin fluctuations and of the local magnetic fields.
The estimated internal fields at zero temperature for the two
different muon-stopping sites are 0.31 and 0.1\,T. 
The smooth transverse muon-spin relaxation rates $\lambda_\mathrm{T}$
in the ordered phase confirm the simple FM structure of
$A_2$La$_2$NiW$_2$O$_{12}$. In both materials, spin fluctuations
are rather weak,  reflected in a small longitudinal muon-spin relaxation rate in both the ferromagnetic- and paramagnetic states. 
In the future, it could be interesting to check if the combined
physical pressure and chemical substitution on the $A$ and $B$' sites can further tune the magnetic competitions in
Sr$_2$La$_2$NiW$_2$O$_{12}$, and eventually lead to magnetic frustration or to a quantum spin-liquid state.

\begin{acknowledgments}
	This work was supported by the Natural Science Foundation of Shanghai 
	(Grants No.\ 21ZR1420500 and 21JC\-140\-2300), Natural Science
	Foundation of Chongqing (Grant No.\ 2022NSCQ-MSX1468), and the Schweizerische 
	Nationalfonds zur F\"{o}r\-der\-ung der Wis\-sen\-schaft\-lichen For\-schung 
	(SNF) (Grants No.\ 200021\_188706 and 206021\_139082). Y.X.\ acknowledges
	support from the Shanghai Pujiang Program (Grant No.\ 21PJ1403100) and the Natural Science Foundation of China (Grant No. 12274125).
\end{acknowledgments}

\bibliography{balaniwo_bib}
\end{document}